\documentclass[12pt]{article}

\usepackage{amssymb}
\usepackage{amsmath}
\usepackage{amsfonts}
\usepackage{epsfig}
\usepackage{anysize}
\marginsize{2cm}{2cm}{2.5cm}{3cm}
\date{}

\usepackage{hyperref}

\def\keywords#1{\par
	\vspace*{8pt}
	{\footnotesize{%\leftskip18pt\rightskip\leftskip
	\noindent{\it Keywords}\/:\ #1\par}}\par}

\def\captionFigureI{%
  \caption{\footnotesize
    (a) The map in Eq.\ (\ref{fixedpoint3}) with $z=2$ and $u=-0.005$ (red), two trajectories are shown with
    $x_0<0$ and $x_0>0$ (blue). The insets show the time dependence of these trajectories that obey
    Eq.\ (\ref{trajectory3}). (b) Rank-order statistics for the occurrence of forest fires (blue dots) from
    Ref. \cite{fires1}. They are reproduced by the smooth curve from Eq.\ (\ref{zipf2}) with $\alpha=2$ and
    $\mathcal{N}^{-1}=-u$ (red). The inset shows the same data (blue) plotted in
    $\ln_{\protect\alpha ^{\prime}}(p_k/p_{\min})$ scale, $p_{k}=1/N(k)$ and $\alpha ^{\prime }=0$.
    The straight line is the corresponding plot of Eq.\ (\ref{expalphaprimepk1}) (red) and evidences the
    extensivity of the entropy in Eq.\ (\ref{microcanonicalentropy1}). See text for details.}
}

\def\captionFigureII{%
\caption{\footnotesize
  (a) Absolute value of trajectory positions $x_t$, $t = 0,1,\ldots$, for
  the logistic map $f_\mu(x)$ at $\mu_\infty$, with initial condition $x_0 = 0$, in
  logarithmic scale as a function of the logarithm of the time $t$, also
  shown by the numbers close to the points. The arrows indicate the distances equivalent to the principal diameters
  $d_{n,0}$. (b) The positions $\left|x_t\right|$ of the trajectory in (a) at selected iteration times $t = 2^{n}-1$
  (black dots) reproduced by Eq.\ (\ref{trajectory4}) with $\alpha=1.7555$ (red). The inset shows the same data plotted
  in $\ln_{\protect\alpha^{\prime }} p_{k}/p_{\min} $ scale, $p_{k}=1/\left|x_{k}\right|$, $p_{\min}=1$ and
  $\alpha ^{\prime }=0.2445$. The straight line evidences the (time) extensivity of the entropy in
  Eq.\ (\ref{microcanonicalentropy1}). See text for description.} 
}

\def\captionFigureIII{%
\caption{\footnotesize
  (a) Iteration time dependence of positions $\theta_t$ in logarithmic scales for the orbit with initial condition
  $\theta_0=0$ at $\Omega_{\infty}^{\prime }$ of the critical circle map $K=1$. The labels indicate
  iteration time $t$, the blue line goes through positions at times of the form $F_{2n}$, while the dotted and dashed
  lines do similarly at times of the form $2F_{2n}$ and $F_{2n}+F_{2n-2}$, respectively. The arrows indicate the
  distances  equivalent to the principal diameters $d_{2n,0}$.
  (b) The positions $\theta_t$ of the trajectory in (a) at selected iteration times $t = F_{2n}-1$ (black dots) reproduced
  by Eq.\ (\ref{trajectory5}) with $\alpha=1.948997$ (red). The inset shows the same data plotted in
  $\ln_{\protect\alpha ^{\prime }}$ scale, $p_{k}=1/\theta_{k}$, $p_{\min}=1$ and $\alpha ^{\prime }=0.051003$.
  The straight line evidences the (time) extensivity of the entropy in Eq.\ (\ref{microcanonicalentropy1}).
  See text for description.} 
}

\begin{document}
\renewcommand{\figurename}{\textbf{\footnotesize Figure}}
\renewcommand{\thefigure}{\mbox{\textbf{\footnotesize\arabic{figure}}}}

\title{\LARGE\bf Entropies for severely contracted configuration space}

\author{G. Cigdem Yalcin\textsuperscript{1}, Carlos Velarde\textsuperscript{2}, Alberto Robledo\textsuperscript{3}\\
\footnotesize 1. Department of Physics, Istanbul University, 34134, Vezneciler, Istanbul, Turkey\\
\footnotesize 2. Instituto de Investigaciones en Matem\'aticas Aplicadas y en Sistemas,\\%
\footnotesize    Universidad Nacional Aut\'onoma de M\'exico\\%
                 %Apartado Postal 70-221, M\'exico 04510 Distrito Federal, M\'exico\\
\footnotesize 3. Instituto de F\'{i}sica y Centro de Ciencias de la Complejidad,\\%
\footnotesize    Universidad Nacional Aut\'onoma de M\'exico,\\%
\footnotesize    Apartado Postal 20-364, M\'exico 01000 DF, Mexico.
}

\maketitle

\abstract{%
 We demonstrate that dual entropy expressions of the Tsallis type apply naturally to statistical-mechanical systems
 that experience an exceptional contraction of their configuration space.  The entropic index $\alpha>1$ describes
 the contraction process, while the dual index $\alpha ^{\prime }=2-\alpha<1$ defines the contraction dimension at
 which extensivity is restored. We study this circumstance along the three routes to chaos in low-dimensional
 nonlinear maps where the attractors at the transitions, between regular and chaotic behavior, drive phase-space
 contraction for ensembles of trajectories. We illustrate this circumstance for properties of systems that find
 descriptions in terms of nonlinear maps. These are size-rank functions, urbanization and similar processes, and
 settings where frequency locking takes place. 
}

%\begin{keyword}
% phase space contraction \sep generalized entropies \sep transitions to chaos
%\end{keyword}
%\keywords{phase space contraction; generalized entropies; transitions to chaos}


\normalsize

\section{Introduction}
 %\label{}

 It is generally acknowledged that the validity of ordinary, Boltzmann-Gibbs (BG), equilibrium statistical
 mechanics rests on the capability of a system composed of many degrees of freedom to transit amongst its many
 possible configurations in a representative manner. The number of configurations of a typical
 statistical-mechanical system increases exponentially with its size, and when these configurations are reachable
 in an adequate fashion through a sufficiently long time period, the indispensable BG properties, ergodicity and
 mixing, are established \cite{uffink1}. Therefore, to explore the limit of validity of BG statistical mechanics it
 is relevant to look at situations where access to configuration space can be controlled to various degrees down to
 a residual set of vanishing measure. A classic example is that of supercooled molecular liquids where glass
 formation signals ergodicity breakdown \cite{debenedetti1}.

Here we refer to an especially tractable family of model systems in which the effect of phase space contraction in
their statistical-mechanical properties can be studied theoretically. These are low-dimensional nonlinear maps that
describe the three different routes to chaos, intermittency, period doublings and quasi periodicity
\cite{schuster1}. Because these systems are dissipative they possess families of attractors, and the dynamics of
ensembles of trajectories towards these attractors constitute realizations of phase space contraction. When the
attractors are chaotic the contraction reaches a limit in which the contracted space has the same dimension as the
initial space, a set of real numbers. But when the attractor is periodic the contraction is extreme and the final
number of accessible configurations is finite. When the attractors at the transitions to chaos are multifractal
sets contraction leads to more involved intermediate cases. Chaotic attractors have ergodic and mixing properties
but those at the transitions to chaos do not \cite{beck1}. We consider them here to discuss their association with
generalized entropies.

The dynamical properties imposed by the attractors at the mentioned transitions to (or out of) chaos in
low-dimensional nonlinear maps can be easily determined \cite{robledo1} and it is our purpose to describe these
properties in terms of phase space contraction. Interestingly, these contractions are found in all cases to be
analytically expressed in terms of the so-called deformed exponential function,
 $\exp_{q}(x)\equiv \left[ 1+(1-q)x\right] ^{1/(1-q)}$, $1\leq q \leq 2$, and these in turn, as we describe below,
 appear associated with dual entropy expressions via the inverse function, the deformed logarithm,
 $\ln_{q}(x)\equiv(1-q)^{-1}[x^{1-q}-1]$. The entropy expressions are of the Tsallis type \cite{tsallis1}, i.e.,
\begin{equation}
      S_{1}[p_{k}]=\sum_{k=0}^{k_{\max}}p_{k}\ln_{\alpha }\,p_{k}^{-1},
      \label{entropy1opt}
\end{equation}%
and%
\begin{equation}
      S_{2}[p_{k}]=-\sum_{k=0}^{k_{\max}}p_{k}\ln_{\alpha ^{\prime }}p_{k},
      \label{entropy2opt}
\end{equation}
where $p_{k}$ are probabilities. The dual entropy expressions satisfy a maximum entropy principle (MEP) and their
values coincide, $S_{2}[p_{k}]=S_{1}[p_{k}]$, when the deformation indexes obey $\alpha ^{\prime }=2-\alpha<1$.
 The dynamics towards the attractor is measured by the index $\alpha$, whereas the index $\alpha ^{\prime }$
 characterizes the contraction achieved by the attractor. Absence of (effective) contraction implies
 $\alpha=\alpha ^{\prime }=1$ and total contraction is signaled by $\alpha=2$ and $\alpha ^{\prime }=0$.
 We define a contraction dimension via the index $\alpha ^{\prime }$,
\begin{equation}
    \alpha ^{\prime }=\frac{\ln \pi_k}{\ln p_k},
    \label{contractdim1}
\end{equation}
where $\pi_k$ and $p_k$ are, respectively, the probabilities associated with the contracted set and the initial set
of configurations.

In the following sections we describe the effect of phase space contraction for the three routes to chaos that
occur in nonlinear maps $f(x)$ of a single variable $x$, $f$ and $x$ real numbers. We consider the effect of
attractors at the transitions to chaos on ensembles of initial conditions that occupy fully one-dimensional
intervals. We begin first with the simplest case of the tangent bifurcation associated with intermittency of type I
\cite{schuster1}, for which the ensemble contracts into a finite set of points. This is the most extreme situation
that leads (in general) to $\alpha=2$ and $\alpha ^{\prime }=0$ \cite{robledo2}, and we corroborate this case with
ranked data for forest fire sizes. Next we consider the accumulation point of period-doubling bifurcations in
quadratic maps, where contraction into the most open region of the multifractal attractor leads to
 $\alpha \simeq 1.7555$ and $\alpha ^{\prime } \simeq 0.2455$ \cite{robledo3}. We refer to systems where period
 doubling is observed or to processes modeled by quadratic maps. Finally, we look at the quasi-periodic transition
 to chaos in the circle map along the golden-mean route and its chosen representative region yields
  $\alpha \simeq 1.9489$ and $\alpha ^{\prime } \simeq 0.0510$ \cite{robledo4}. We describe this contraction in
  terms of its mode-locking property \cite{schuster1} widely observed elsewhere.

\section{Tangent bifurcation}
 %\label{}

A common account of the tangent bifurcation, that mediates the transition between a chaotic attractor and an
attractor of period $n$, starts with the composition $f^{(n)}(x)$ of a one-dimensional map $f(x)$, e.g. the logistic map,
at such bifurcation, followed by an expansion around the neighborhood of one of the $n$ points tangent to the line with
unit slope \cite{schuster1}. That is 
\begin{equation}
x^{\prime }=f^{(n)}(x)=x+ux^{z}+\cdots,\;x\geq 0,\;z>1,  \label{tangent1}
\end{equation}%
where $x^{z}\equiv sign(x)\left\vert x\right\vert ^{z}$. The functional composition Renormalization Group (RG) fixed-point
map is the solution $f^{\ast }(x)$ of%
\begin{equation}
f^{\ast }(f^{\ast }(x))=\lambda ^{-1}f^{\ast }(\lambda x)
\label{fixedpoint1}
\end{equation}%
together with a specific value for $\lambda $ that upon expansion around $x=0 $ reproduces Eq.\ (\ref{tangent1}). An
exact analytical expression for $x^{\prime }=f^{\ast }(x)$ was obtained long ago \cite{schuster1}. This is 
\begin{equation}
x^{\prime 1-z}=x^{1-z}+(1-z)u  \label{fixedpoint2}
\end{equation}%
or, equivalently,
\begin{equation}
x^{\prime }=x\exp_{z}(ux^{z-1}),  \label{fixedpoint3}
\end{equation}%
with $\lambda =2^{1/(z-1)}$. Repeated iteration of
Eq.\ (\ref{fixedpoint2}) leads to% 
\begin{equation}
x_{t}^{1-z}=x_{0}^{1-z}+(1-z)ut  \label{trajectory1}
\end{equation}%
or%
\begin{equation}
\ln_{z}x_{t}=\ln_{z}x_{0}+ut.  \label{trajectory2}
\end{equation}%
So that the iteration number or time $t$ dependence of all trajectories is
given by%
\begin{equation}
x_{t}=x_{0}\exp_{z}\left[ x_{0}^{z-1}ut\right] ,  \label{trajectory3}
\end{equation}%
where the $x_{0}$ are the initial positions. In Fig.\ 1a we plot the map in Eq.\ (\ref{fixedpoint3}) when $z=2$ together
with two trajectories initiated at $x_{0}<1$ and $x_{0}>1$, that in the insets are shown as functions of $t$ (also
reproduced by Eq.\ (\ref{trajectory3})). The $q$-deformed properties of the tangent bifurcation are discussed at greater
length in Ref. \cite{robledo2}. An ensemble of trajectories with initial conditions $x_{0}$ distributed within an
interval $X \leq x_{0} <0$, $X<0$ arbitrary, undergoes progressive phase space contraction ending up into the point
$x=0$.  

\begin{figure*}[tb]\scriptsize
  \vspace*{.05in}
  \centering
  \includegraphics[width=0.9\textwidth]{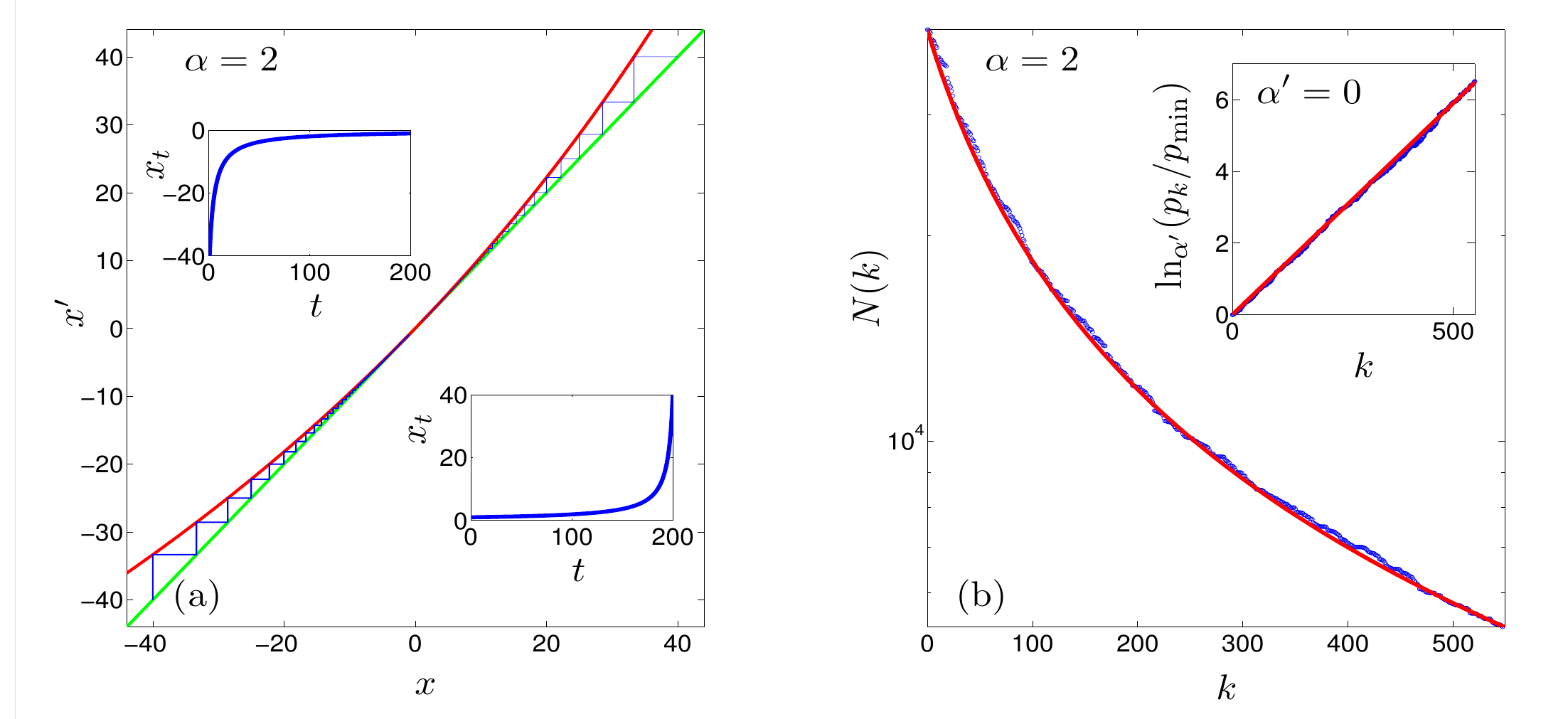}
  \parbox{0.9\textwidth}{\captionFigureI}
\end{figure*}

Interestingly, a manifestation of the dynamical properties at the tangent bifurcation appears in ranked data that follow
Zipf's law for large enough rank. It has been shown \cite{robledo5}, \cite{robledo6} that size-rank functions, as given
by a basic argument \cite{pietronero1} that generalizes Zipf's law, are strictly analogous to the RG fixed-point map
trajectories in Eq.\ (\ref{trajectory3}). The expression for the size-rank function $N(k)$, the magnitude of the data $N$
for rank $k$,  

\begin{equation}
N(k)=N_{\max}\exp_{\alpha }[-N_{\max}^{\alpha -1}\mathcal{N}^{-1}k],
\label{zipf2}
\end{equation}%
where $\mathcal{N}=k_{\max}$ is the total number of data, becomes that in Eq.\ (\ref{trajectory3}) with the
identifications  $k=t$, $\mathcal{N}^{-1}=-u$, $N(k)=-x_{t}$, $N_{\max}=-x_{0}$ and $\alpha =z$.
We note that the most common value for the degree of nonlinearity at tangency is $z=2$, obtained when the map is
analytic at $x=0$ with nonzero second derivative, and this
implies $\alpha =2$, close to the values observed for many sets of real data, as this conforms with the classical Zipf's
law form $N(k)\sim k^{-1}$ when $k$ is large. The inverse of $N(k)$, $p_{k}=1/N(k)$, the (uniform) probability for the
occurrence of each unit that constitutes $N(k)$, is given by 
\begin{equation}
p_{k}=p_{\min }\exp_{\alpha ^{\prime }}(p_{\min }^{\alpha ^{\prime }
-1}\mathcal{N}^{-1}k),  \label{expalphaprimepk1}
\end{equation}
where $p_{\min }=1/N_{\max}$. In Fig.\ 1b we plot $N(k)$ in Eq.\ (\ref{zipf2}) when $\alpha=2$ together with ranked data
of forest fires areas \cite{fires1} while in the inset we plot Eq.\ (\ref{expalphaprimepk1}) when $\alpha ^{\prime }=0$
also with the corresponding forest fire data. We notice that, if $W(k_{\max}) \equiv N_{\max}/{N(k_{\max})}$ is
identified as the number of configurations of the system of size equal to the maximum rank $k_{\max}$ that generates
the data set $N(k), k=0,1,\ldots,k_{\max}$, then from Eq.\ (\ref{expalphaprimepk1}) we observe that the size-dependent
entropy \cite{robledo7} 
\begin{equation}
S(k_{\max})\equiv\ln_{\alpha ^{\prime }}W(k_{\max}),
\label{microcanonicalentropy1}
\end{equation}%
is extensive.

\section{Period-doubling accumulation point}
 %\label{}
   
The classic example of functional composition RG fixed-point map is the solution of Eq.\ (\ref{fixedpoint1}) associated
with the period-doubling accumulation points of unimodal maps \cite{schuster1}. In practice it is often illustrated by
use of the quadratic $z=2$ logistic map $f_{\mu }(x)=1-\mu x^{2}$, $-1\leq x\leq 1$, $0\leq \mu \leq 2$, with the
control parameter located at $\mu =\mu_{\infty }(z=2)=1.401155189092\cdots$, the value for the accumulation point of
the main period-doubling cascade \cite{beck1}. The scaling factor, known as Feigenbaum's universal constant, is
$\lambda(z=2)=-2.50290\cdots$ (for convenience we denote below its absolute value with the same symbol).    

Fig.\ 2a shows two features of the trajectory at $\mu_{\infty }(z=2)$ with
initial condition at $x_{0}=0$ that are relevant to our discussion. Notice
that in this figure we have used absolute values of iterated positions $\left| x_{t}\right| $
to facilitate the use logarithmic scales. The labels correspond to iteration
times $t$. The first visible feature in the figure is that the positions fall within equally-spaced horizontal bands.
Since all positions visited at odd times form the top band one half of the attractor lies there.
Inspection of the iteration times for positions within the subsequent bands indicates that one quarter of the
attractor forms the second band, one eighth the third band, and so on. Repeated functional composition of
the unimodal map eliminates bands successively starting with the top band, and these in turn
can be recovered (approximately) by repeated rescaling by a factor equal to $\lambda $.
This removal and recuperation of bands correspond to a graphical construction of the functional
composition and rescaling of the RG transformation.

\begin{figure*}[tb]
  \centering
  \includegraphics[width=.9\textwidth]{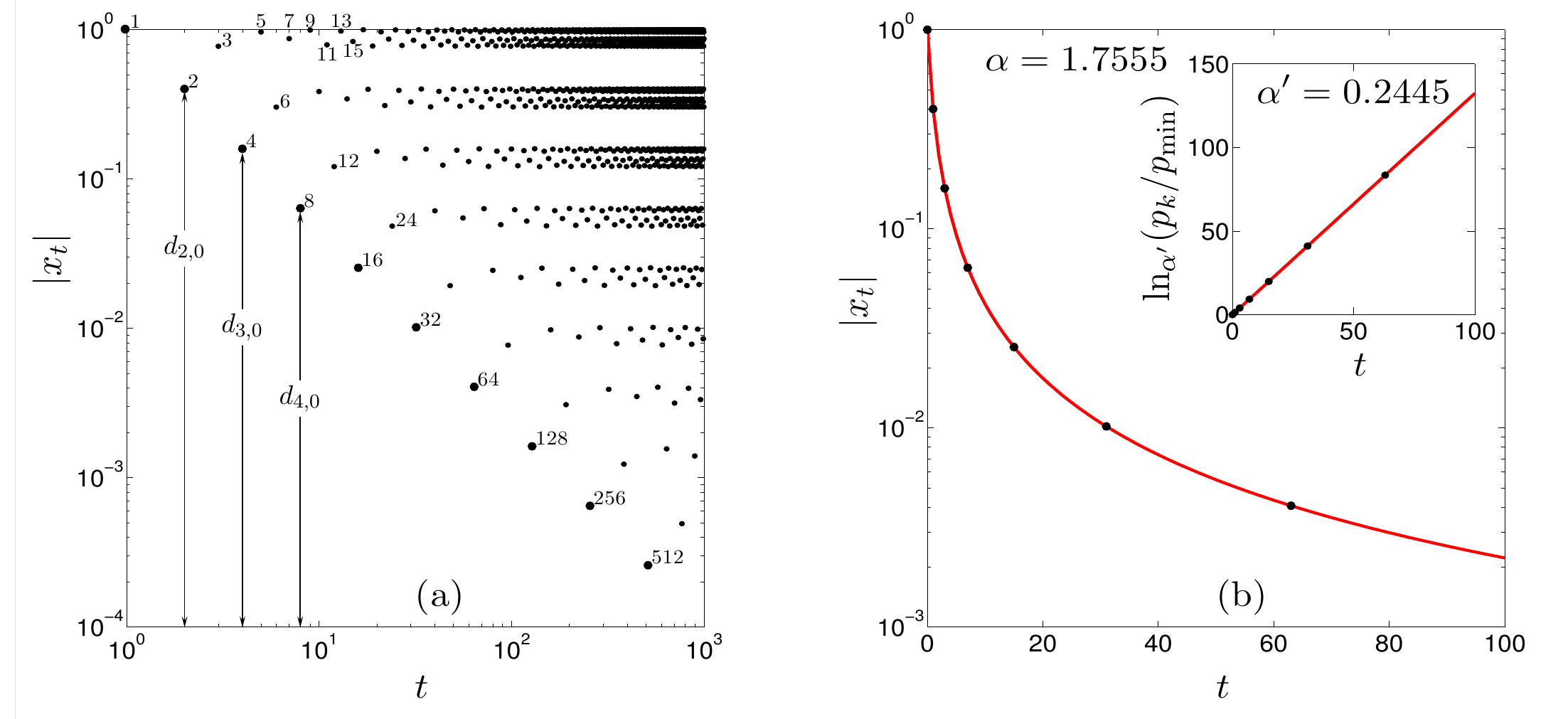}
  \parbox{0.9\textwidth}{\captionFigureII}
\end{figure*}

The second feature evident in Fig.\ 2a is that all the attractor positions fall into a well-defined family
of straight diagonal lines, all with the same slope. All the attractor positions can
be allocated into subsequences formed by the time subsequences $t=(2k+1)2^{n}-1$, with
$n=0,1,2,\ldots$ and fixed $k=0,1,2,\ldots$. Thus, the subsequences $\left| x_{t}\right|$ have each a common power-law decay 
$\left| x_{t}\right|\simeq (t+1) ^{1/1-\alpha}$, with $\alpha=1+\ln 2/\ln \lambda (z)$,
$\alpha\simeq 1.7555$ when $z=2$ \cite{robledo1}, \cite{robledo3}. Interestingly, these subsequences
can be seen \cite{robledo1}, \cite{robledo3} to reproduce  the positions of the so-called
`superstable' periodic trajectories \cite{schuster1}. In particular, the positions for the main
subsequence $k=0$ are given by $\left| x_{2^{n-1}}\right| \simeq d_{n,0}=$ $\lambda^{-n}$,
where $d_{n,0}\equiv \left| f_{\overline{\mu }_{n}}^{(2^{n-1})}(0)\right| $ is the `$n$-th principal
diameter' \cite{schuster1}. With use of $\lambda^{-n}\equiv(1+t)^{-\ln\lambda/\ln2}$,
 $t = 2^{n}-1$, this subsequence can be expressed as 
\begin{equation}
  \left| x_{t}\right| =\exp_{\alpha}(-\Lambda_{\alpha}t)
\label{trajectory4}
\end{equation}
with $\alpha$ as above and $\Lambda_{\alpha}=(z-1)\ln \lambda /\ln 2$.
See \cite{robledo1}, \cite{robledo3}. 

The band structure in Fig.\ 2a involves families of phase-space gaps that decrease in width as power laws
(equal sizes in logarithmic scales of the figure). These gaps are formed sequentially, beginning with the largest
one, in the dynamics towards the attractor at $\mu_{\infty }$. This process has a hierarchical organization and
can be observed explicitly by placing an ensemble of initial conditions $x_{0}$ distributed uniformly across
phase space and record their positions at subsequent times \cite{robledo1, robledo8, robledo9}. The main gaps in Fig.\ 2a
decrease with the same power law of the principal diameters $d_{n,0}$ described above and expressed as the deformed
exponential in Eq.\ (\ref{trajectory4}). The locations of this specific family of consecutive gaps advance monotonically
toward the sparsest region of the multifractal attractor located at $x=0$ \cite{robledo1, robledo3, robledo8}. The decreasing
values of $\left| x_{t}\right|$, $t = 2^{n}-1$ in Eq.\ (\ref{trajectory4}) with increasing $n$, describe phase-space contraction
along iteration time evolution as these intervals represent the widths of the gaps forming consecutively, the diameter $d_{n,0}$
being the width of the gap formed after $t = 2^{n+1}-1$ \cite{robledo8}. Similarly other families of diameters represent widths
of gaps leading to the multifractal attractor. In Fig.\ 2b we plot Eq.\ (\ref{trajectory4}) with $\alpha\simeq 1.7555$ that
reproduces the positions $x_t$ of the trajectory initiated at $x_{0}$ for iteration times $t = 2^{n}-1$, $n=0,1,2,\ldots$. In the
inset we plot $\ln_{\alpha^{\prime }} p_k$, $p_{k}=1/\left|x_{k}\right|$ with $\alpha ^{\prime }=2-\alpha =0.2445$. The
straight line corresponds to Eq.\ (\ref{microcanonicalentropy1}) (with $p_{\min}=1$ and $\mathcal{N}=\ln2/\ln\lambda)$ in
Eq.\ (\ref{expalphaprimepk1}), and corroborates the (time) extensivity of entropy.

\section{Golden-mean route to chaos} 
  %\label{}

The quasi periodic route to chaos is often studied by means of the circle map,%
\begin{equation}
f_{\Omega ,K}(\theta )=\theta +\Omega -K/(2\pi) \sin 2\pi \theta ,\;\textnormal{mod}\;1,
  \label{circle1}
\end{equation}%
where the control parameters $\Omega $\ and $K$ are, respectively, the bare winding number
and the degree of nonlinearity \cite{schuster1}. Another quantity relevant to the dynamics generated by this map is the
dressed winding number $\omega\equiv \lim_{t\rightarrow \infty }[\theta_t-\theta_0]/t$. Locked motion (a periodic attractor)
occurs when $\omega$ is rational and unlocked motion (a quasi periodic attractor) when $\omega$ is irrational. We are
interested in the critical circle map $K=1$ when locked motion occurs for all $\Omega $, $0\leq \Omega \leq 1$, except
for a multifractal set of unlocked values. Sequences of locked motion values of $\Omega$ can be used to select attractors of
increasing periods such that a transition to chaos is obtained at their infinite-period (quasi periodic) accumulation points \cite{schuster1}.

A well-known specific case of the above is the sequence of rational approximations to the reciprocal of the golden mean
$\omega_{\rm gm}=(\sqrt{5}-1)/2\simeq 0.618034$. This sequence is formed by the winding
numbers $\omega_{n}=F_{n-1}/F_{n}$, where $F_{n}$ are the Fibonacci numbers $F_{n+1}=$ $F_{n}+F_{n-1}$.
The route to chaos is the family of attractors with increasing periods $F_{n}$, $n\rightarrow \infty $. Amongst these attractors
it is possible to select specific values of $\Omega $ with the superstable property \cite{schuster1}. As before,
a superstable trajectory of period $T$ satisfies $df^{(T)}(\theta_{0})/d\theta =0$, and is one that contains as
one of its positions $\theta =0$. There are two superstable families of trajectories, the first at control parameter
values $\Omega_{n}$, $n=1,2,\ldots$ with winding numbers $\omega_{n}=F_{n-1}/F_{n}$ and accumulation point at
 $\Omega_{\infty }\simeq 0.606661$ which corresponds to $\omega_{\rm gm}$.  The second family with winding numbers
$\omega_{n}^{\prime }=F_{n-2}/F_{n}$, with $\Omega_{n}^{\prime }$, $n=1,2,\ldots$ and accumulation point at
$\Omega_{\infty }^{\prime }=1-\Omega_{\infty }\simeq 0.393339$ which corresponds to
 $\omega_{\rm gm}^{2}\simeq\allowbreak 0.381966$ \cite{robledo4}. 

As with the period-doubling route, the quasiperiodic route to chaos displays universal scaling properties. And an RG
approach, analogous to that for the tangent bifurcation and the period doubling cascade, has been carried out for the
critical circle map \cite{schuster1}. The fixed-point map $f^{\ast }(\theta )$ of an RG transformation that consists of
functional composition and rescaling appropriate for maps with a zero-slope cubic inflection point (like the critical circle map) satisfies 
\begin{equation}
f^{\ast }(\theta )=\lambda_{\rm gm}f^{\ast }(\lambda_{\rm gm}f^{\ast }(\theta /\lambda_{\rm gm}^{2})),
\label{fixed-point1}
\end{equation}%
where (for the golden mean route) $\lambda_{\rm gm}\simeq -1.288575$ is a universal constant
\cite{schuster1}. (We denote below its absolute value with the same symbol). This constant describes the scaling of the
distance $d_{n,0} $ from $\theta =0$ to the nearest element of the orbit with $\omega_{n}$. These are the distances
analogous to the principal diameters in the previous section \cite{schuster1}.

For our purposes we refer only to one family of winding numbers, $\omega_{n}^{\prime }=F_{n-2}/F_{n}$.
Fig.\ 3a shows the trajectory at $\Omega =\Omega_{\infty }^{\prime }$ starting at $\theta_0=0$ in logarithmic scales
where the labels indicate iteration times $t$. Similarly to Fig.\ 2a, a conspicuous feature in Fig.\ 3a is
that positions fall along straight diagonal lines, again, a signal of multiple power law behavior. Notice that the
positions of the main diagonal in Fig.\ 3a correspond to the times $F_{2n}$, $n=1,2,3,\ldots$ The succeeding diagonals
above it appear grouped together (see Ref. \cite{robledo4} for a description).
Also similarly to Fig.\ 2a, in Fig.\ 3a we show the positions for the main subsequence
that constitutes the lower bound of the entire trajectory. These positions are
identified to be $\theta_{F_{2n}} \simeq d_{2n,0}=\Omega_{\infty }^{\prime }\lambda_{\rm gm}^{-2n}$,
where $d_{2n,0}$ is the `$2n$-th principal diameter' defined at the $F_{2n}$-supercycle,
the distance of the orbit position nearest to $\theta=0$ \cite{schuster1}, \cite{robledo4}.
With use of
$\lambda_{\rm gm}^{-2n}\equiv(1+t)^{2\ln\lambda_{\rm gm}/\ln\omega_{\rm gm}}$,
$t = F_{2n}-1$, the main subsequence $\theta_{F_{2n}}$ can be expressed as 
\begin{equation}
\theta_{t} =\Omega_{\infty }^{\prime }\exp_{\alpha}(-\Lambda_{\alpha}t),
\label{trajectory5}
\end{equation}
with $\alpha=1-(1/2)\ln\omega_{\rm gm} /\ln\lambda_{\rm gm}\simeq 1.948997$ and
$\Lambda_{\alpha}=-2\ln \lambda_{\rm gm} / \ln \omega_{\rm gm}$. See \cite{robledo1}, \cite{robledo4}.

As in the case of period doublings, the dynamics towards the attractor at
$\Omega_{\infty }^{\prime }$ of an ensemble of trajectories (with say, uniformly
distributed initial conditions in $0\leq \theta\leq 1$) successively forms gaps of 
decreasing lengths in phase space (the interval $0\leq \theta\leq 1$). The gaps
have a hierarchical structure that ends up at the multifractal set of the attractor
positions partially shown in Fig.\ 3a. A fragment of this process can be observed
through the decreasing lengths of the diameters $d_{2n,0}$ described above and
expressed as the deformed exponential in Eq.\ (\ref{trajectory5}). The locations of
this specific family of consecutive gaps advance monotonically toward the sparsest
region of the multifractal attractor located at $\theta=0$ \cite{robledo4}. As in the previous section, the decreasing
lengths of the principal diameters $d_{2n,0}$ with increasing $n$, equal to the values of $\left|\theta_{t}\right|$,
 $t = F_{2n}-1$, in Eq.\ (\ref{trajectory5}), describe phase-space contraction via formation of successive gaps. Similarly
other families of diameters represent widths of gaps leading to the multifractal attractor. In Fig.\ 3b we plot
Eq.\ (\ref{trajectory5}) with $\alpha\simeq 1.948997$ that reproduces the positions $\theta_t$ of the trajectory
initiated at $\theta_{0}$ for iteration times $t = F_{2n}-1$, $n=0,1,2,\ldots$. In the inset we plot $p_k$ in
$\ln_{\alpha ^{\prime }}$ scale, $p_{k}=1/\theta_{k}$ with $\alpha ^{\prime }=2-\alpha =0.051003$. The straight line
corresponds to Eq.\ (\ref{microcanonicalentropy1}) (with $p_{min}=1/\Omega_{\infty}^{\prime }$ and
$\mathcal{N}={-\Omega_{\infty}^{\prime }}^{1-\alpha^{\prime}}\ln\omega_{\rm gm}/2\ln\lambda_{\rm gm})$
in Eq.\ (\ref{expalphaprimepk1}), and corroborates the (time) extensivity of entropy.

The interpretation of phase space contraction is that as iteration time advances when
 $\Omega=\Omega_{\infty }^{\prime}$ the trajectory positions in the interval $0<\theta_t<\theta_{F_{2n}}$
 approach but fail to attain mode-locking for
dressed winding number $\omega_{n}^{\prime }=F_{2n-2}/F_{2n}$ at $t = F_{2(n+1)}-1$, $n=0,1,2,\ldots$. These intervals
shrink according to Eq.\ (\ref{trajectory5}) as $n\rightarrow \infty$ and in this limit there is no mode-locking, only
quasi periodic motion with $\omega=\omega_{\rm gm}^2$.

\begin{figure*}[bt]
  \centering
  \includegraphics[width=.9\textwidth]{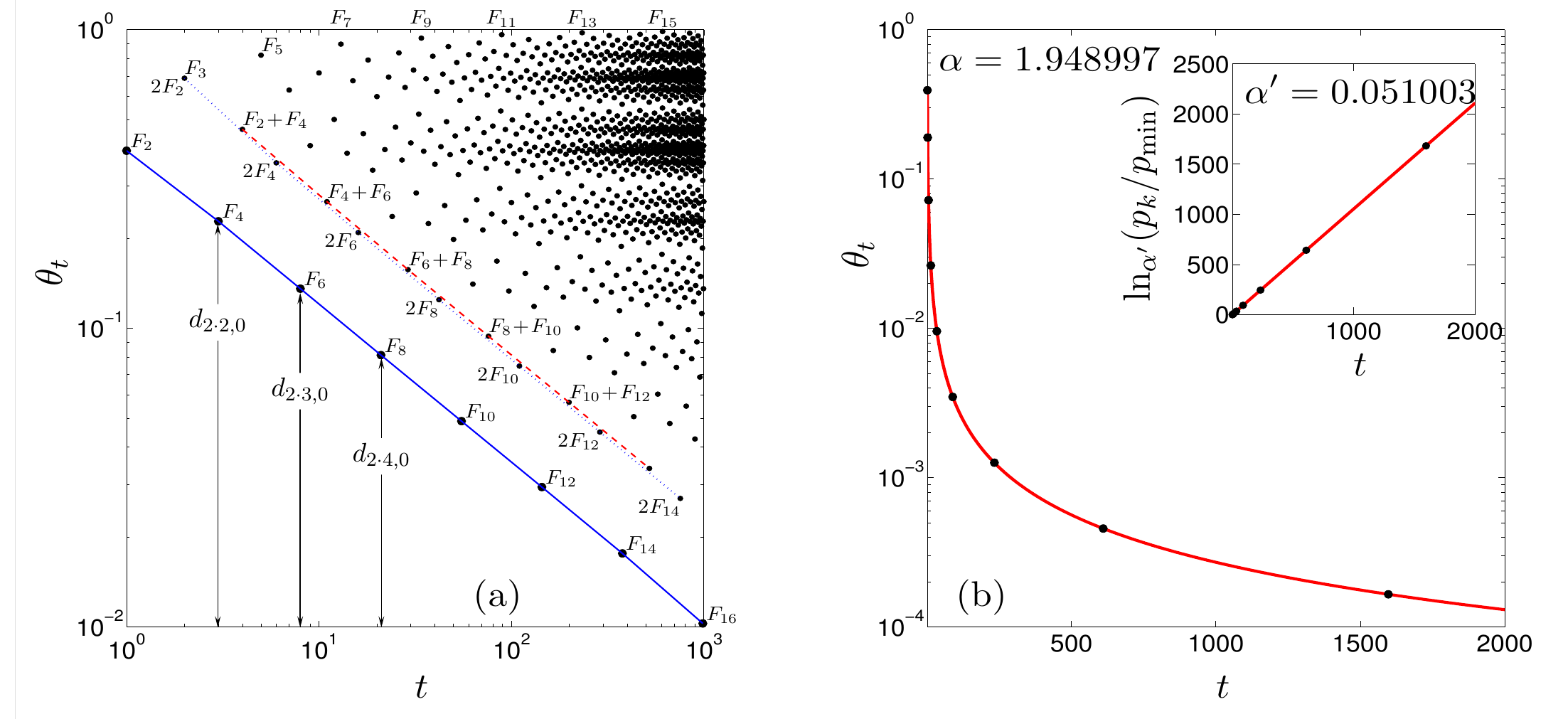}
  \parbox{0.9\textwidth}{\captionFigureIII}
\end{figure*}

\section{Phase space contraction and maximum entropy}
 %\label{}

As we have seen in the previous sections the contraction of phase space guided by the attractors at the three types of
transitions to chaos is described quantitatively by the time evolution of trajectory positions. The expressions for
these trajectory positions are conveniently obtained from the RG fixed-point maps at the transitions to chaos, and in
all cases are exactly given by $q$-exponential functions. See Eqs. (\ref{trajectory3}), (\ref{trajectory4}) and
(\ref{trajectory5}). These functions replace the ordinary Boltzmann weights in generalized statistical-mechanical
expressions associated with entropies of the Tsallis type that are written in terms of the inverse function, the
$q$-logarithm. See Eqs. (\ref{entropy1opt}) and (\ref{entropy2opt}).  We show now that these expressions can be obtained
also from a Maximum Entropy Principle (MEP) and discuss further the occurrence of the dual indexes $\alpha$ and
 $\alpha^{\prime }=2-\alpha<1$ in relation to phase-space contraction. 

Consider the entropy functional $\Phi_{1}[p_{k}]$ of the probabilities $p_{k}$, $k=0,1,2,\ldots,k_{\max}$ with
Lagrange  multipliers $a$ and $b$,%
\begin{equation}
\Phi_{1}[p_{k}]=S_{1}[p_{k}]+a\left[ \sum_{k=0}^{k_{\max}}p_{k}-\mathcal{P}%
\right] +b\left[ \sum_{k=0}^{k_{\max}}kp_{k}-\mathcal{K}\right] ,
\label{functional1}
\end{equation}%
where the entropy expression $S_{1}[p_{k}]$ has the trace form
\begin{equation}
S_{1}[p_{k}]=\sum_{k=0}^{k_{\max}}s_{1}(p_{k}).  \label{entropy1}
\end{equation}%
Optimization via $\partial \Phi_{1}[p_{k}]/\partial p_{k}=0$, $k=0,1,2,\ldots,k_{\max}$,
 gives $s_{1}^{\prime }(p_{k})=-a-bk$.
Now, the choices $s'_1(p_k) = \ln_\alpha p_k^{-1}  - p_{\min}^{-(1-\alpha)} + (1-\alpha)\mathcal{N}^{-1} k$,
$a = -\ln_\alpha p_{\min}^{-1} + p_{\min}^{-(1-\alpha)}$, $b=\alpha \mathcal{N}^{-1}$, lead to%
\begin{equation}
\ln_{\alpha }\ p_{k}^{-1}=\ln_{\alpha }p_{\min }^{-1}-\mathcal{N}^{-1}k.
\label{logalphapk1}
\end{equation}%
or%
\begin{equation}
p_{k}^{-1}=p_{\min }^{-1}\exp_{\alpha }(-p_{\min }^{1-\alpha }\mathcal{N}%
^{-1}k).  \label{expalphapk1}
\end{equation}%
from which we immediately recover Eq.\ (\ref{zipf2}). But also, importantly, $S_1[p_k]$ in Eq (\ref{entropy1}) becomes
Eq.\ (\ref{entropy1opt}). 

Consider next the same functional as above but with the probabilities $p_{k}$ replaced now by the new set
 $\pi_{k}$, $k=0,1,2,\ldots,k_{\max}$. We write
\begin{equation}
\Phi_{2}[\pi_{k}]=S_{2}[\pi_{k}]+c\left[ \sum_{k=0}^{k_{\max}}\pi_{k}-\mathcal{P} ^{\prime }%
\right] +d\left[ \sum_{k=0}^{k_{\max}}k\pi_{k}-\mathcal{K} ^{\prime }\right] .
\label{functional2}
\end{equation}
So that $s'_2(\pi_k) = - c-dk$, with the choices
$s'_2(\pi_k) = - \ln_{\alpha'} \pi_k  - \pi_{\min}^{1-{\alpha'}} - (1-{\alpha'})\mathcal{N}^{-1} k$, 
$c=\pi_{\min}^{1-{\alpha'}}+\ln_{\alpha'}\pi_{\min}$,
$d=(2-\alpha')\mathcal{N}^{-1}$,  
lead to%
\begin{equation}
\ln_{\alpha ^{\prime } }\ \pi_{k}=\ln_{\alpha ^{\prime }  }\pi_{\min }+\mathcal{N}^{-1}k.
\label{logalphapk2}
\end{equation}%
or%
\begin{equation}
\pi_{k}=\pi_{\min }\exp_{\alpha ^{\prime } }(\pi_{\min }^{\alpha ^{\prime }-1 }\mathcal{N}%
^{-1}k),  \label{expalphapk2}
\end{equation}%
which, also importantly, when used to evaluate $S_2[\pi_k]$ leads to Eq.\ (\ref{entropy2opt}).

If the probabilities $p_k$ and $\pi_k$ correspond, respectively, to the initial and the contracted sets of
configurations,
and if they are both normalizable in their own spaces, then to recover one from the other we require a
relationship such as
\begin{equation}
\pi_{k}=p_{k}^{\alpha ^{\prime }},
\label{contractdim2}
\end{equation}
with $\alpha ^{\prime }<1$ when the contracted set has a vanishing measure with respect to the initial one. Therefore we
note that in the MEP procedure it is not necessary to make use of the constraint
\begin{equation}
\sum_{k=0}^{k_{\max}}k\ p_{k}^{\alpha ^{\prime }}=\textnormal{constant},
\label{constraint1}
\end{equation}
commonly used in the derivation of generalized entropies \cite{tsallis1}, including Ref. \cite{robledo7} (where there
appears some non-consequential faux pas). Instead, the distinction between the initial and contracted space of
configurations indicates the introduction of a contraction dimension $\alpha ^{\prime }$ via Eqs. (\ref{contractdim1})
or (\ref{contractdim2}).    

\section{Discussion}
 %\label{}

We have discussed the association of dual entropy expressions of the Tsallis type with the dynamical properties of
attractors in low-dimensional iterated maps along the three routes to chaos: intermittency, period doublings and
quasi-periodicity. The attractors at the transitions to chaos provide a natural mechanism by means of which ensembles of
trajectories are forced out of almost all phase space positions and become confined into a finite or (multi)fractal set
of permissible positions. Such drastic contraction of phase space leads to nonergodic and nonmixing dynamics that is
described by the dual entropic indexes $\alpha>1$ and $\alpha ^{\prime }=2-\alpha<1$. The first fixes the deformation of
the exponential that measures the degree of contraction along (iteration) time evolution, and the second defines a
contraction dimension, cf.\ Eq.\ (\ref{contractdim1}), such that extensivity of entropy is restored. The dual entropy
expressions are compatible with the same maximum entropy principle. When the contraction of phase space leads to a set
of configurations of the same measure as the original phase space, e.g., an interval or finite collection of intervals
of real numbers, one has $\alpha=\alpha ^{\prime }=1$, the entropy expressions are the same and maintain the usual BG
expression. 
        
We chose to examine the statistical-mechanical effect of configuration space contraction at the renowned transitions to
chaos in low-dimensional nonlinear maps, as these are perhaps the simplest situations where ergodicity and mixing
properties breakdown. But in their own, the properties of these model systems manifest in natural phenomena. There are
abundant examples of ranked data that obey (approximately) the empirical Zipf power law and these have been shown to
comply with the tangent bifurcation property \cite{robledo7}. The period doubling cascade to chaos has been considered
recently in many model systems ranging from fluid convection \cite{yahata1} to urbanization processes
\cite{urbanization1}. The basic features of the quasi periodic route to chaos have been famously measured in forced
Rayleigh-Benard convection \cite{libchaber1} and in periodically perturbed cardiac cells \cite{glass1}.   

\section{Acknowledgements}
The authors are grateful for the interest of Murray Gell-Mann in this work, and for his valuable advice and
recommendations. G.C.Y. gratefully acknowledges the hospitality of the Santa Fe Institute.
 G.C.Y. was supported by the Scientific Research Projects Coordination Unit of Istanbul University
 with project number 49338.
% C.V. acknowledges support provided by IIMAS-UNAM.
 A.R. acknowledges support by DGAPA-UNAM-IN103814 and CONACyT-CB-2011-167978 (Mexican Agencies).


%\bibliographystyle{ieeetr}
%\bibliography{arXiv.bib}

\end{document}